\documentclass[aps,prb,twocolumn,groupedaddress,showpacs]{revtex4}
\usepackage{graphicx}

\begin{document}


\title{Lattice dynamics and the electron-phonon interaction in Ca$_2$RuO$_4$}



\author{H. Rho,$^{1,2}$ S. L. Cooper,$^2$
S. Nakatsuji,$^{3}$ H. Fukazawa,$^3$ and Y. Maeno,$^{3,4}$ }

\affiliation{ $^{1}$Department of Physics, Chonbuk National
University, Jeonju 561-756, Korea\\
$^{2}$Department of Physics and Frederick Seitz Materials Research
Laboratory,
University of Illinois at Urbana-Champaign, Urbana, Illinois 61801\\
$^{3}$Department of Physics, Kyoto University, Kyoto 606-8502,
Japan\\
$^{4}$International Innovation Center, Kyoto University, Kyoto
606-8501, Japan
 }


\date{\today}

\begin{abstract}

We present a Raman scattering study of Ca$_2$RuO$_4$, in which we
investigate the temperature-dependence of the lattice dynamics and
the electron-phonon interaction below the metal-insulator
transition temperature ({\it T}$_{\rm MI}$).  Raman spectra
obtained in a backscattering geometry with light polarized in the
ab-plane reveal 9\,B$_{1g}$ phonon modes (140, 215, 265, 269, 292,
388, 459, 534, and 683 cm$^{-1}$) and 9\,A$_g$ phonon modes (126,
192, 204, 251, 304, 322, 356, 395, and 607 cm$^{-1}$) for the
orthorhombic crystal structure (Pbca$-$D$_{2h}^{15}$).  With
increasing temperature toward {\it T}$_{\rm MI}$, the observed
phonon modes shift to lower energies and exhibit reduced spectral
weights, reflecting structural changes associated with the
elongation of the RuO$_6$ octahedra. Interestingly, the phonons
exhibit significant increases in linewidths and asymmetries for
{\it T} $>$ {\it T}$_{\rm N}$.  These results indicate that there
is an increase in the effective number of electrons and the
electron-phonon interaction strengths as the temperature is raised
through {\it T}$_{\rm N}$, suggesting the presence of orbital
fluctuations in the temperature regime {\it T}$_{\rm N}$ $<$ {\it
T} $<$ {\it T}$_{\rm MI}$.

\end{abstract}


\maketitle


{\section{INTRODUCTION}}

Like many other transition-metal oxides that exhibit strong
correlations among spin, charge, orbital, and lattice degrees of
freedom, Ca$_{2-x}$Sr$_x$RuO$_4$ exhibits many exotic phenomena
throughout its rich phase diagram. \cite{Maeno, Nakatsuji1, Hotta,
Nakatsuji2, Anisimov, Lee, Jung, Mizokawa1, Snow, Rho, Nakatsuji3}
For instance, not only do the ruthenates exhibit orbital ordering,
\cite{Anisimov, Lee, Jung, Mizokawa1}, but also orbital-dependent
superconductivity, heavy-mass Fermi liquid behavior, and
metal-insulator transitions. \cite {Maeno, Nakatsuji1, Nakatsuji2}

The single-layer ruthenate material Sr$_2$RuO$_4$ is a
superconductor below {\it T}$_{\rm c}$ = 1.5 K, possibly with an
unconventional {\it p}-wave pairing state. \cite{Ishida}
Substitution of Ca for Sr significantly distorts the lattice
structure and lowers the crystal symmetry from cubic to
orthorhombic, giving rise to remarkable changes in magnetic,
electronic, and structural properties. \cite{Nakatsuji1,
Nakatsuji2, Friedt}  The ground state of Ca$_2$RuO$_4$ is
antiferromagnetic (AF) insulating. \cite{Satoru, Alexander}  With
increasing temperature, Ca$_2$RuO$_4$ becomes paramagnetic (PM)
insulating at {\it T}$_{\rm N}$ = 113 K and then PM metallic at
{\it T}$_{\rm MI}$ = 357 K. \cite{Nakatsuji1, Alexander, Fukazawa}
The metal-insulator (MI) transition is first-order, as evidenced
by the observation of thermal hysteresis.  Further, the transition
is driven by an elongation of the RuO$_6$ octahedra with
increasing temperature through {\it T}$_{\rm MI}$, and therefore
the insulating and metallic states are characterized by short
(S-Pbca) and long (L-Pbca) c-axis lattice parameters,
respectively. \cite{Friedt, Braden} As a function of
Sr-substitution, {\it x}, the AF ground state of
Ca$_{2-x}$Sr$_x$RuO$_4$ persists for {\it x} $<$ 0.2. Sr
substitution changes both {\it T}$_{\rm N}$ and {\it T}$_{\rm
MI}$, and significantly affects magnetic, electronic, orbital, and
structural correlations in this material. For instance, Raman
scattering results obtained on Ca$_{2-x}$Sr$_x$RuO$_4$ have shown
that Sr substitution causes a dramatic increase in the
renormalization of the two-magnon (2M) energy and linewidth, and
an increase of the electron-phonon interaction strength.
\cite{Rho} Raman scattering measurements further suggest that the
exchange coupling constant {\it J} in Ca$_{2-x}$Sr$_x$RuO$_4$ is
relatively insensitive to pressure \cite{Snow} and Sr content.
\cite{Rho}

Other complex oxides are known to exhibit significant changes in
phonon dynamics near important phase transitions.  For example,
La$_{0.7}$Ca$_{0.3}$MnO$_3$ displays a significant shift of the
transverse optical phonon energy near the MI transition, giving
rise to an increased electron-phonon interaction related to
changes in the lattice parameters. \cite{Kim}  Raman studies of
LaMnO$_3$ reveal an effect of the spin-lattice interaction near
the transition to AF order. \cite{Podobedov}  In
La$_{2-x}$Sr$_x$CuO$_4$, an increase of Sr concentration causes
the appearance of an asymmetric phonon lineshape and a dramatic
decrease of the phonon intensity associated with the apical oxygen
vibration, indicating that the electron-phonon interaction is also
important in this material. \cite{Nimori}  Inelastic neutron
scattering and x-ray scattering studies have suggested the
presence of orbital fluctuations in LaTiO$_3$. \cite{Keimer} As
described in Ref. 21, an electronic continuum and anomalous phonon
behavior with a Fano profile observed in the Raman response of
{\it R}TiO$_3$ ({\it R}=rare earth) \cite{Reedyk} may also
indicate the presence of orbital fluctuations in the insulating
region of this system. Therefore, it is of great interest to study
how spin, charge, orbital, and lattice correlations evolve in
Ca$_2$RuO$_4$ through different phase transitions, particularly as
a means of comparing the exotic properties of this system to those
of complex oxides such as the cuprates, manganites, and titanates.
In this paper, we use the unique strengths of Raman scattering to
explore lattice dynamics and the electron-phonon interaction in
Ca$_2$RuO$_4$ as a function of temperature between 10 K and 300 K.

\begin{table*}
\caption{\label{tab:table2}Site symmetries and IR's of the atoms
in Ca$_2$RuO$_4$ with space group Pbca$-$D$_{2h}^{15}$. Mode
classifications are: $\Gamma$$_{\rm Raman}$ = 9 (A$_g$ + B$_{1g}$
+ B$_{2g}$ + B$_{3g}$), $\Gamma$$_{\rm infrared}$ = 11 (B$_{1u}$ +
B$_{2u}$ + B$_{3u}$), $\Gamma$$_{\rm silent}$ = 12 A$_u$, and
$\Gamma$$_{\rm acoustic}$ = B$_{1u}$ + B$_{2u}$ + B$_{3u}$.  The
corresponding polarization tensor elements for each of the
Raman-active factor group species are: A$_g$ $\rightarrow$
$\alpha$$_{xx}$, $\alpha$$_{yy}$, $\alpha$$_{zz}$; B$_{1g}$
$\rightarrow$ $\alpha$$_{xy}$, $\alpha$$_{yx}$; B$_{2g}$
$\rightarrow$ $\alpha$$_{xz}$, $\alpha$$_{zx}$; and B$_{3g}$
$\rightarrow$ $\alpha$$_{yz}$, $\alpha$$_{zy}$.}
\begin{ruledtabular}
\begin{tabular}{ccccc}
&Atom&Site symmetry&IR's&\\
\hline &Ru& C$_i$ & 3 (A$_u$ +B$_{1u}$ + B$_{2u}$ + B$_{3u}$)& \\
&Ca& C$_1$ & 3 (A$_g$ + B$_{1g}$ + B$_{2g}$ + B$_{3g}$ + A$_u$ +B$_{1u}$ + B$_{2u}$ + B$_{3u}$)& \\
&O(1)& C$_1$ & 3 (A$_g$ + B$_{1g}$ + B$_{2g}$ + B$_{3g}$ + A$_u$ +B$_{1u}$ + B$_{2u}$ + B$_{3u}$)& \\
&O(2)& C$_1$ & 3 (A$_g$ + B$_{1g}$ + B$_{2g}$ + B$_{3g}$ + A$_u$ +B$_{1u}$ + B$_{2u}$ + B$_{3u}$)& \\
\end{tabular}
\end{ruledtabular}
\end{table*}

$\newline$
{\section{EXPERIMENT}}

A single-crystal sample of Ca$_2$RuO$_4$, which was grown by a
floating-zone method, \cite{Nakatsuji1, Fukazawa, Nakatsuji4} was
mounted inside a continuous He-flow cryostat.  The 647.1-nm
excitation wavelength from a Kr-ion laser was used in a
backscattering geometry with the propagation vector ({\bf k})
oriented along the c axis of the sample, {\bf k} $\parallel$
c-axis.  Scattered light from the sample was dispersed using a
triple-stage spectrometer, and then recorded using a
liquid-nitrogen-cooled charge-coupled device (CCD) detector.
Various polarization configurations of the incident and scattered
light were employed in order to identify the scattering symmetries
of the Raman spectra obtained for Ca$_2$RuO$_4$: ({\bf E}$_i$,
{\bf E}$_s$) = (x, y), B$_{1g}$ symmetry; ({\bf E}$_i$, {\bf
E}$_s$) = (x, x), A$_g$ symmetry; and ({\bf E}$_i$, {\bf E}$_s$) =
(x$'$, x$'$), B$_{1g}$ + A$_g$ symmetry, where {\bf E}$_i$ and
{\bf E}$_s$ are the incident and the scattered polarization
directions, respectively, B$_{1g}$ and A$_g$ are irreducible
representations (IR's) of the space group D$_{2h}$, and x
$\parallel$ [1,0,0], y $\parallel$ [0,1,0], and x$'$ $\parallel$
[1,1,0].  All the Raman spectra were corrected, first, by removing
the CCD dark current response, and second, by normalizing the
spectrometer response using a calibrated white light source.
Finally, the corrected spectra were divided by the Bose thermal
factor, giving rise to the spectral responses displayed in this
paper. These responses are proportional to the imaginary part of
the Raman susceptibility.

$\newline$
{\section{ RESULTS AND DISCUSSION }}

\begin{figure}
\centering
\includegraphics[width=9cm]{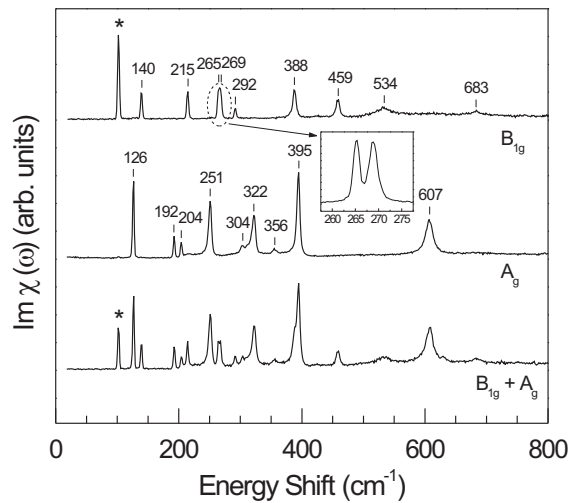}
\vspace{0.1cm}
\caption{\label{}Polarized Raman spectra at {\it T}
= 10 K with B$_{1g}$, A$_g$, and B$_{1g}$ + A$_g$ scattering
symmetries from top to bottom, respectively.  The inset shows a
high-resolution Raman spectrum, indicating two resolved 265 and
269 cm$^{-1}$ phonon modes.}
\end{figure}

Ca$_2$RuO$_4$ has an orthorhombic crystal structure (space group
Pbca$-$D$_{2h}^{15}$) with four formula units per unit cell.  A
factor-group analysis, summarized in Table I, yields a total of 81
$\Gamma$-point phonons, of which 36 [9 (A$_g$ + B$_{1g}$ +
B$_{2g}$ + B$_{3g}$)] are Raman-active modes involving Ca,
in-plane oxygen [O(1)], and apical oxygen [O(2)] ions, 33
[11(B$_{1u}$ + B$_{2u}$ + B$_{3u}$)] are infrared-active modes,
and 12 (12 A$_u$) are silent modes.  The Ru ions are located at a
center of inversion symmetry and, thus, do not participate in any
Raman-active phonon modes.  Unlike tetragonal Sr$_2$RuO$_4$, in
which 2\,A$_{1g}$ and 2\,E$_g$ symmetry optical phonons are
Raman-active, orthorhombic Ca$_2$RuO$_4$ exhibits numerous phonon
lines.  This reflects the fact that substitution of Ca for Sr
strongly distorts the RuO$_6$ octahedra, causing a rotation of the
octahedra around the c axis, and a tilt of the octahedra around an
axis on the RuO$_2$ plane. \cite {Friedt, Braden}  As shown in
Fig.\,1, polarized Raman spectra in a backscattering geometry
(with the propagation vector {\bf k} $\parallel$ c-axis) reveal
all of the phonon modes corresponding to each of scattering
symmetries: 9\,B$_{1g}$ symmetry modes in ({\bf E}$_i$, {\bf
E}$_s$) = (x, y), 9\,A$_g$ symmetry modes in ({\bf E}$_i$, {\bf
E}$_s$) = (x, x), and 9\,B$_{1g}$ + 9\,A$_g$ symmetry modes in
({\bf E}$_i$, {\bf E}$_s$) = (x$'$, x$'$) polarization
configurations, respectively.  Note in the inset of Fig. 1 that
the B$_{1g}$ phonon peak energies assigned at 265 and 269
cm$^{-1}$ are clearly resolved in a high-resolution Raman
spectrum.  More specific assignments of the observed optical
phonons to particular atomic normal modes will require lattice
dynamic calculations.

\begin{figure}
\centering
\includegraphics[width=8.5cm]{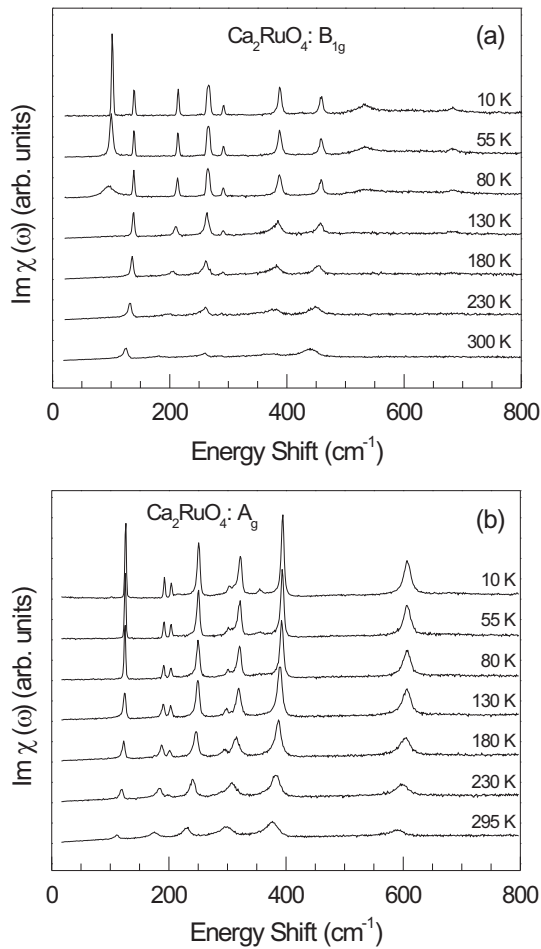}
\caption{\label{}(a) B$_{1g}$ and (b) A$_g$ Raman
spectra with increasing temperature from 10 K to room
temperature.}
\end{figure}

Figure 1 also shows a Raman-active mode at 102 cm$^{-1}$ (denoted
with an asterisk) that is observed only in the B$_{1g}$ scattering
geometry. This mode is likely associated with a 2M scattering
response, although we cannot completely rule out the possibility
that it is a one-magnon excitation.  The 2M scattering response,
which involves a photon-induced flipping of spins on
nearest-neighbor Ru sites, provides useful information concerning
the AF correlations. \cite{Fleury, Sugai, Rho, Snow}  Using the
fact that the 2M energy for an S\,=\,1 AF insulator is given by
$\hbar$$\omega$\,=\,6.7\,{\it J},\cite{Sugai} where {\it J} is the
in-plane exchange coupling constant between nearest-neighbor
Ru-4$d^4$ sites, we can estimate {\it J} = 15.2 cm$^{-1}$ in
Ca$_2$RuO$_4$.  With increasing temperature toward {\it T}$_{\rm
N}$, as shown in Fig. 2(a), the 2M response weakens in intensity,
broadens in linewidth, and shifts to lower energy, reflecting the
reduction of the AF correlations in Ca$_2$RuO$_4$ as the
temperature is increased to {\it T}$_{\rm N}$.  Unlike the
cuprates, the 2M scattering intensity in Ca$_2$RuO$_4$ diminishes
rapidly above {\it T}$_{\rm N}$, indicating that local AF order
disappears for {\it T} $>$ {\it T}$_{\rm N}$.  More details of the
2M characteristics in Ca$_{2-x}$Sr$_x$RuO$_4$ have been described
elsewhere, including the effects of pressure \cite {Snow} and Sr
substitution. \cite {Rho}

With increasing temperature toward {\it T}$_{\rm MI}$, the
out-of-plane Ru-O(2) bond length is nearly unchanged for {\it T}
$<$ {\it T}$_{\rm N}$, but gradually elongates as temperature is
raised above {\it T}$_{\rm N}$. \cite{Friedt, Braden}  In order to
elucidate the temperature-dependence of the lattice parameters and
the electron-phonon interaction below {\it T}$_{\rm MI}$, both
B$_{1g}$ and A$_g$ symmetry Raman spectra from Ca$_2$RuO$_4$ were
studied as a function of increasing temperature from 10 to 300 K,
as summarized in Figs. 2(a) and 2(b).  There are several key
features observed in the Raman spectra as a function of increasing
temperature, including (i) a softening of all the B$_{1g}$ and the
A$_g$ optical phonon energies, (ii) a decrease of phonon spectral
weights, and (iii) a significant broadening and increased
asymmetry of phonon lineshapes across {\it T}$_{\rm N}$.

\begin{figure}
\centering
\includegraphics[width=9cm]{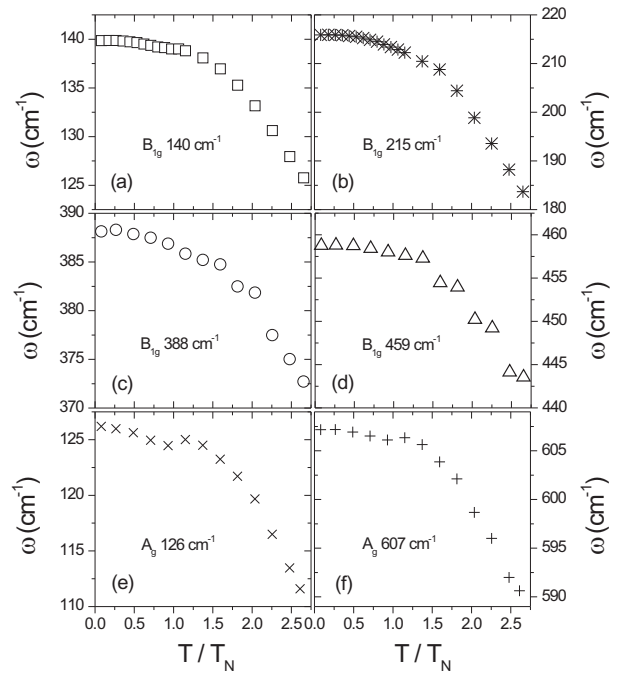}
\caption{\label{}Temperature-dependent frequency
shifts of B$_{1g}$ modes at (a) 140, (b) 215, (c) 388, (d) 459
cm$^{-1}$, and of A$_g$ modes at (e) 126 and (f) 607 cm$^{-1}$.}
\end{figure}

The systematic shifts of phonon peaks to lower energies with
increasing temperature through {\it T}$_{\rm N}$ primarily reflect
an elongation of the RuO$_6$ octahedra along the c axis.  Figures
3(a) to 3(f) summarize the phonon energy changes with increasing
temperature for some representative B$_{1g}$ (140, 215, 388, and
459 cm$^{-1}$) and A$_g$ (126 and 607 cm$^{-1}$) optical phonon
modes. The phonon-energy shifts are negligible for {\it T} $<$
{\it T}$_{\rm N}$, indicating little or no change in the lattice
parameters in this temperature regime.  In contrast, remarkable
phonon-energy shifts are observed for {\it T} $>$ {\it T}$_{\rm
N}$.  Most of phonon modes display downward shifts of $\sim$ 15
cm$^{-1}$ as temperature is raised from 10 to 300 K.
Interestingly, the B$_{1g}$ phonon at 215 cm$^{-1}$ shows a much
more dramatic energy shift of $\sim$ 30 cm$^{-1}$.  These Raman
results are consistent with neutron scattering measurements, which
show that there is little change in the lattice parameters for
{\it T} $<$ {\it T}$_{\rm N}$, but that there is a significant
change in the lattice parameters for {\it T}$_{\rm N}$ $<$ {\it T}
$\sim$ {\it T}$_{\rm MI}$. \cite{Friedt, Braden}  All the other
B$_{1g}$ and A$_g$ optical phonons decrease in energy with
increasing temperature.  Note that all the B$_{1g}$ phonon modes
exhibit a dramatic decrease in intensity as temperature is raised
toward {\it T}$_{\rm MI}$ ($\sim$ 357 K $\sim$ 3.2{\it T}$_{\rm
N}$), possibly reflecting increased damping of the modes by
thermally activated carriers.

One notes in Figs. 2(a) and 2(b) that the B$_{1g}$ and the A$_g$
phonon lineshapes at low temperatures for {\it T} $<$ {\it
T}$_{\rm N}$ are quite symmetric and narrow.  In contrast, with
increasing temperature through {\it T}$_{\rm N}$, the phonon
linewidths broaden significantly and the lineshapes become
increasingly asymmetric.  The latter reveals a Fano effect, caused
by the interaction between the discrete phonon state and a broad
electronic continuum of states. \cite{Fano}  Similar behavior has
been observed in Raman spectra of numerous other strongly
correlated materials such as Ca$_{2-x}$Sr$_x$RuO$_4$, \cite{Rho}
La$_{1-x}$Ca$_x$MnO$_3$, \cite{Naler} Ca$_3$Ru$_2$O$_7$,
\cite{Liu} Sr$_2$RuO$_4$, \cite{Sakita} and {\it R}TiO$_{3}$ ({\it
R} = rare earth). \cite{Reedyk}

To study the temperature-dependence of the electron-phonon
interaction in Ca$_2$RuO$_4$ in detail, the temperature-dependence
of the phonon linewidths and asymmetries of the B$_{1g}$ and A$_g$
phonon modes were extracted by fitting these modes to a Fano
lineshape, $I$($\omega$) = $I$$_0$($q$ + $\epsilon$)$^2$/(1 +
$\epsilon$$^2$), where $\epsilon$ = ($\omega$ $-$
$\omega$$_0$)/$\Gamma$, $\omega$$_0$ is the phonon energy,
$\Gamma$ is the effective phonon linewidth, and $q$ is the
asymmetry parameter.  In this way, one obtains information on the
electron-phonon interaction.  The inverse of the asymmetry
parameter, $1/|q|$, is proportional to the electron-phonon
coupling strength V and the imaginary part of the electronic
susceptibility $\rho$ according to $1/q$ $\sim$ $V\rho$.
\cite{Rho, Fano, Naler}  Moreover, the electron-phonon coupling
contribution to the phonon linewidth can be estimated from the
fractional change in the phonon damping rate below {\it T}$_{\rm
MI}$, [$\Gamma$($T$) $-$ $\Gamma$($RT$)]/$\omega$$_0$ =
$\Delta$$\Gamma$/$\omega$$_0$ $\propto$ N(0)$\omega$$_0$$\lambda$,
where N(0) is the electronic density of states at the Fermi
surface, $\omega$$_0$ is the phonon energy, and $\lambda$ is the
dimensionless electron-phonon coupling parameter.  \cite{Nyhus,
Allen, Axe}  The parameter $\lambda$ is related to the BCS
parameter N(0)V$_{ph}$, where V$_{ph}$ is the pairing potential
arising from the electron-phonon interaction. \cite{Allen, Axe,
McMillan}  Therefore, by carefully monitoring the phonon
linewidths, as well as the inverse asymmetry parameters, as a
function of temperature, one can obtain useful information
regarding the evolution of the electronic density of states and
the electron-phonon coupling strengths in Ca$_2$RuO$_4$.

\begin{figure}
\centering
\includegraphics[width=8.5cm]{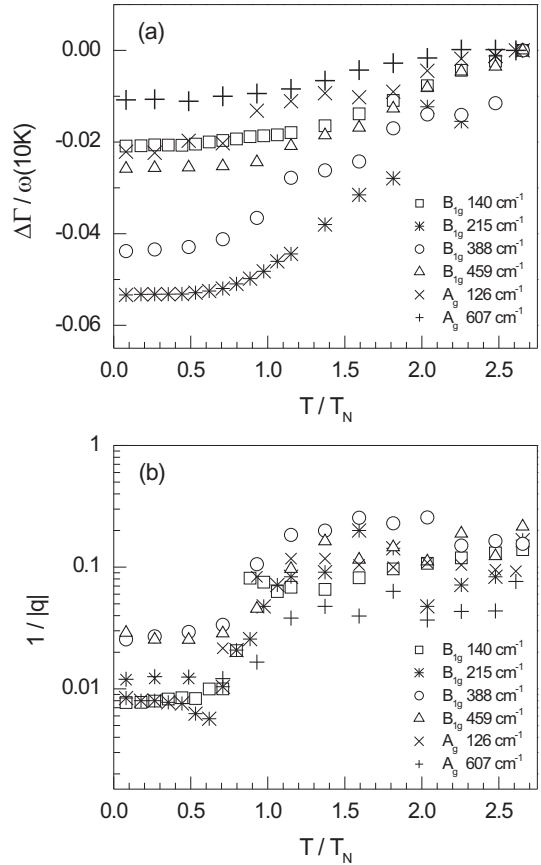}
\caption{\label{}(a) Phonon linewidth changes
divided by the corresponding phonon energy at {\it T} = 10 K,
[$\Gamma$($T$) $-$ $\Gamma$($RT$)]/$\omega$(10 K) =
$\Delta$$\Gamma$/$\omega$(10 K), as a function of temperature
normalized to {\it T}$_{\rm N}$. (b) Magnitudes of inverse of the
asymmetry parameters, $1/|q|$, as a function of temperature
normalized to {\it T}$_{\rm N}$.}
\end{figure}

The role of the electronic contribution to the system above {\it
T}$_{\rm N}$, which influences the substantial phonon energy
renormalizations observed above {\it T}$_{\rm N}$, can be explored
by plotting as a function of temperature the phonon linewidth
changes divided by the corresponding phonon energy at {\it T} = 10
K, $\Delta$$\Gamma$/$\omega$(10\,K).  These plots are displayed in
Fig.\,4(a).  Interestingly, in contrast to the negligible change
in phonon linewidths observed below {\it T}$_{\rm N}$, there is
significant broadening in the phonon linewidths above {\it
T}$_{\rm N}$.  For example, the phonon linewidths at 300 K are
significantly broader than those at 10 K,
$\Delta$$\Gamma$/$\omega$$_0$ $\sim$ 5.3 $\%$, which is even
larger than the fractional broadening observed in the correlation
gap material FeSi, $\Delta$$\Gamma$/$\omega$$_0$ $\sim$ 3.5 $\%$.
\cite{Nyhus}  We attribute the systematic broadening of the phonon
linewidths with increasing temperature above {\it T}$_{\rm N}$ to
an increase of the effective number of electrons in the system for
{\it T} $>$ {\it T}$_{\rm N}$. Indeed, Jung et al. recently
reported that the effective number of electrons systematically
increases, and the optical gap closes, with increasing temperature
above {\it T}$_{\rm N}$ in Ca$_2$RuO$_4$. \cite{Jung}

The evolution of the electron-phonon interaction in Ca$_2$RuO$_4$
can be carefully illustrated by plotting the inverse of the
asymmetry parameter, $1/|q|$, for different modes as a function of
temperature, as shown in Fig.\,4(b).  Note that the magnitudes of
the inverse asymmetry parameters, $1/|q|$, which are obtained from
the representative B$_{1g}$ and the A$_g$ phonon modes, are
negligible at low temperatures, but increase significantly as
temperature is raised through T$_N$. Even when the temperature is
much lower than the MI transition temperature, there is no
additional increase of the inverse asymmetry parameters for {\it
T}/{\it T}$_{\rm N}$ $>$ $\sim$ 1.2.  These results strongly
suggest that the largest increase in the electron-phonon coupling
strength occurs near {\it T}$_{\rm N}$, rather than near {\it
T}$_{\rm MI}$. By contrast, between {\it x} = 0.45 and 0.76 in
La$_{1-x}$Ca$_x$MnO$_3$, the $1/|q|$ value decreases linearly with
increasing {\it x} in the ferromagnetic low-temperature metallic
region ({\it x} $\leq$ 0.52), and vanishes in the AF region (for
{\it x} $>$ 0.52). \cite{Naler}  Note also that the magnitudes of
the inverse asymmetry parameters, $1/|q|$, in the PM insulating
region of Ca$_2$RuO$_4$ for {\it T}$_{\rm N}$ $<$ {\it T} $<$ {\it
T}$_{\rm MI}$ are comparable to those in the metallic region of
Sr$_2$RuO$_4$ (Ref. 29) and La$_{1-x}$Ca$_x$MnO$_3$. \cite{Naler}

It is interesting to note that previous Raman results on
Ca$_{2-x}$Sr$_x$RuO$_4$ have revealed that substitution of Sr for
Ca increases the electron-phonon coupling strength, as evidenced
by an increase of both the inverse asymmetry parameters and the
phonon linewidths with Sr substitution at {\it T} = 10 K.
\cite{Rho} Moreover, for the Sr-substituted samples ({\it x} =
0.06 and 0.09), a charge gap observed at 10 K was found to close
well below the MI transition temperature, suggesting that the
intermediate temperature regime between {\it T}$_{\rm N}$ and {\it
T}$_{\rm MI}$ consists of a coexistence of insulating S-Pbca and
metallic L-Pbca phases. \cite{Rho}  In contrast, in this study,
the broadening of phonon linewidths and the increase of the
$1/|q|$ values upon heating through {\it T}$_{\rm N}$ ($\ll$ {\it
T}$_{\rm MI}$) in Ca$_2$RuO$_4$ are probably not attributable to
the coexistence of metallic and insulating phases in this
temperature regime for several reasons, including the high
stoichiometry of the sample, and the absence of any residual low
frequency conductivity in Ca$_2$RuO$_4$ in the temperature regime
{\it T}$_{\rm N}$ $<$ {\it T} $<$ {\it T}$_{\rm MI}$. \cite {Lee,
Jung}

In Ca$_2$RuO$_4$, the observed increase in both the
electron-phonon coupling strength and the number of effective
carriers with increasing temperature near {\it T}$_{\rm N}$
strongly suggests that enhanced orbital fluctuations, which are
associated with increased electron transfer between the {\it
d}$_{\rm xy}$ and {\it d}$_{\rm yz/zx}$ orbitals, are responsible
for the behavior observed in this temperature regime. Indeed, a
recent O 1{\it s} x-ray absorption spectroscopy study of
Ca$_{2-x}$Sr$_x$RuO$_4$ ({\it x} = 0.0 and 0.09) has suggested
that orbital fluctuations gradually increase upon heating even in
the insulating region well below {\it T}$_{\rm MI}$.
\cite{Mizokawa}  Note that the inverse asymmetry parameters and
the spectral linewidths for the phonon mode near 300 cm$^{-1}$ in
the insulating {\it R}TiO$_3$ ({\it R} = Gd, Sm, Nd, Pr, Ce, La)
exhibit the largest values in LaTiO$_3$, \cite{Reedyk} suggesting
that orbital fluctuations are important in LaTiO$_3$, as pointed
out by Keimer et al. \cite{Keimer}

$\newline$
{\section{ CONCLUSIONS }}

In summary, temperature-dependent Raman spectra of Ca$_2$RuO$_4$
allow us to explore the lattice dynamics near the MI transition
temperature of this system.  With increasing temperature through
the N\'{e}el temperature, the B$_{1g}$ and the A$_g$ phonon modes
exhibit a substantial shift to lower energies.  Moreover, the
phonons significantly broaden and exhibit increasingly asymmetric
lineshapes upon heating in the vicinity of {\it T}$_{\rm N}$.
These results demonstrate that both the electron-phonon coupling
strength and the effective number of electrons increase as
temperature is raised through {\it T}$_{\rm N}$, suggesting that
orbital fluctuations are present in the PM insulating region in
the temperature regime {\it T}$_{\rm N}$ $<$ {\it T} $<$ {\it
T}$_{\rm MI}$.

$\newline$
\begin{acknowledgments}

Work in Korea was supported by Korea Research Foundation Grant
(KRF-2004-005-C00003).  Work in Illinois was supported by the
National Science Foundation under Grant No. DMR02-44502 and by the
Department of Energy through the Materials Research Laboratory
under Grant No. DEFG02-91ER45439.  The work at Kyoto Univ. was
supported in part by Grants-in-Aid for Scientific Research from
JSPS and for the 21st Century COE ``Center for Diversity and
Universality in Physics" from MEXT of Japan.

\end{acknowledgments}



\bibliography{basename of .bib file}

\begin{thebibliography}{}


\bibitem{Maeno} Y. Maeno, H. Hashimoto, K. Yoshida, S. Nishizaki, T. Fujita, J.G. Bednorz, and F. Lichtenberg, Nature (London) \textbf{372}, 532 (1994).

\bibitem{Nakatsuji1} S. Nakatsuji and Y. Maeno, Phys. Rev. Lett. \textbf{84}, 2666 (2000); S. Nakatsuji and Y. Maeno, Phys. Rev. B \textbf{62}, 6458 (2000).

\bibitem{Hotta} T. Hotta and E. Dagotto, Phys. Rev. Lett. \textbf{88}, 17201 (2002).

\bibitem{Nakatsuji2} S. Nakatsuji, D. Hall, L. Balicas, Z. Fisk, K. Sugahara, M. Yoshioka, and Y. Maeno, Phys. Rev. Lett. \textbf{90}, 137202 (2003).

\bibitem{Anisimov} V.I. Anisimov, I.A. Nekrasov, D.E. Kondakov, T.M. Rice, and M. Sigrist, Eur. Phys. J. B \textbf{25}, 191 (2002).

\bibitem{Lee} J.S. Lee, Y.S. Lee, T.W. Noh, S.-J. Oh, J. Yu, S. Nakatsuji, H. Fukazawa, and Y. Maeno, Phys. Rev. Lett. \textbf{89}, 257402 (2002).

\bibitem{Jung} J.H. Jung, Z. Fang, J.P. He, Y. Kaneko, Y. Okimoto, and Y. Tokura, Phys. Rev. Lett. \textbf{91}, 56403 (2003).

\bibitem{Mizokawa1} T. Mizokawa, L. H. Tjeng, G. A. Sawatzky, G. Ghiringhelli, O.
Tjernberg, N. B. Brookes, H. Fukazawa, S. Nakatsuji, and Y. Maeno,
Phys. Rev. Lett. \textbf{87}, 77202 (2001).

\bibitem{Snow} C.S. Snow, S. L. Cooper, G. Cao, J.E. Crow, H. Fukazawa, S. Nakatsuji, and Y. Maeno, Phys. Rev. Lett. \textbf{89}, 226401 (2002).

\bibitem{Rho} H. Rho, S.L. Cooper, S Nakatsuji, H. Fukazawa, and Y. Maeno, Phys. Rev. B \textbf{68}, 100404(R) (2003).

\bibitem{Nakatsuji3} S. Nakatsuji, V. Dobrosavljevi\'{c}, D. Tanaskovi\'{c}, M. Minakata, H. Fukazawa, and Y. Maeno, Phys. Rev. Lett.
\textbf{93}, 146401 (2004).

\bibitem{Ishida} K. Ishida, H. Mukuda, Y. Kitaoka, Z.Q. Mao, H. Fukazawa, and Y. Maeno, Phys. Rev. B \textbf{63}, 60507(R) (2001).

\bibitem{Friedt} O. Friedt, M. Braden, G. Andr\'{e}, P. Adelmann, S. Nakatsuji, and Y. Maeno, Phys. Rev. B \textbf{63}, 174432 (2001).

\bibitem{Satoru} S. Nakatsuji, S. Ikeda, and Y. Maeno, J. Phys. Soc. Jpn. \textbf{66}, 1868 (1997).

\bibitem{Alexander} C.S. Alexander, G. Cao, V. Dobrosavljevic, S. McCall, J.E. Crow, E. Lochner, and R.P. Guertin, Phys. Rev. B \textbf{60}, 8422(R) (1999).

\bibitem{Fukazawa} H. Fukazawa, S. Nakatsuji, and Y. Maeno, Physica B \textbf{281$\&$282}, 613 (2000).

\bibitem{Braden} M. Braden, G. Andr\'{e}, S. Nakatsuji, and Y. Maeno, Phys. Rev. B \textbf{58}, 847 (1998).

\bibitem{Kim} K.H. Kim, J.Y. Gu, H.S. Choi, G.W. Park, and T.W. Noh, Phys. Rev. Lett. \textbf{77}, 1877 (1996).

\bibitem{Podobedov} V.B. Podobedov, A. Weber, D.B. Romero, J.P. Rice, and H.D. Drew, Phys. Rev. B \textbf{58}, 43 (1998).

\bibitem{Nimori} S. Nimori, S. Sakita, F. Nakamura, T. Fujita, H. Hata, N. Ogita, and M. Udagawa, Phys. Rev. B \textbf{62}, 4142 (2000).

\bibitem{Keimer} B. Keimer, D. Casa, A. Ivanov, J.W. Lynn, M.v. Zimmermann, J.P.
Hill, D. Gibbs, Y. Taguchi, and Y. Tokura, Phys. Rev. Lett.
\textbf{85}, 3946 (2000).

\bibitem{Reedyk} M. Reedyk, D.A. Crandle, M. Cardona, J.D. Garrett, and J.E. Greedan, Phys. Rev. B \textbf{55}, 1442 (1997).

\bibitem{Nakatsuji4} S. Nakatsuji and Y. Maeno, J. Solid State Chem. \textbf{156}, 26 (2001).

\bibitem{Fleury} P.A. Fleury and R. Loudon, Phys. Rev. \textbf{166}, 514 (1968); P.A. Fleury and H.J. Guggenheim, Phys. Rev. Lett. \textbf{24}, 1346 (1970).

\bibitem{Sugai} S. Sugai, M. Sato, T. Kobayashi, J. Akimitsu, T. Ito, H. Takagi, S. Uchida, S. Hosoya, T. Kajitani, and T. Fukuda, Phys. Rev. B \textbf{42}, 1045 (1990).

\bibitem{Fano} U. Fano, Phys. Rev. \textbf{124}, 1866 (1961).

\bibitem{Naler} S. Naler, M. Rubhausen, S. Yoon, S.L. Cooper, K. H. Kim, and S.W. Cheong, Phys. Rev. B \textbf{65}, 92401 (2002).

\bibitem{Liu} H.L. Liu, S. Yoon, S.L. Cooper, G. Cao, and J.E. Crow, Phys. Rev. B \textbf{60}, R6980 (1999).

\bibitem{Sakita} S. Sakita, S. Nimori, Z.Q. Mao, Y. Maeno, N. Ogita, and M. Udagawa, Phys. Rev. B \textbf{63}, 134520 (2001).

\bibitem{Nyhus} P. Nyhus, S.L. Cooper, and Z. Fisk, Phys. Rev. B \textbf{51}, 15626 (1995).

\bibitem{Allen} P.B. Allen, Phys. Rev. B \textbf{6}, 2577 (1972).

\bibitem{Axe} J.D. Axe and G. Shirane, Phys. Rev. Lett. \textbf{30}, 214 (1973).

\bibitem{McMillan} W.L. McMillan, Phys. Rev. \textbf{167}, 331 (1968).

\bibitem{Mizokawa} T. Mizokawa, L.H. Tjeng, H.-J. Lin, C.T. Chen, S. Schuppler, S.
Nakatsuji, H. Fukazawa, and Y. Maeno, Phys. Rev. B \textbf{69},
132410 (2004).

\end{thebibliography}

\end{document}